# Coalescing Neutron Stars as Gamma Ray Bursters ?

M. Ruffert[1], H.-Th. Janka[1,2], W. Keil[1], and G. Schäfer[3]

[1]Max-Planck-Institut für Astrophysik, Postfach 1523, D-85740 Garching
[2]Dept. of Astron. & Astrophys., Univ. of Chicago, 5640 S. Ellis Ave., Chicago, IL 60637
[3]Arbeitsgruppe "Gravitationstheorie", Friedrich-Schiller Universität, D-07743 Jena

We investigate the dynamics and evolution of coalescing neutron stars. The three-dimensional Newtonian equations of hydrodynamics are integrated by the 'Piecewise Parabolic Method' (PPM; Colella & Woodward, 1984). The grid is Cartesian and equidistant with a resolution of $64^3$ or $128^3$, which allows us to calculate the self-gravity via fast Fourier transforms. Although the code is purely Newtonian, we do include the effects of the emission of gravitational waves on the hydrodynamics in a way originally proposed by Blanchet, Damour & Schäfer (1990). In particular we treat the gravitational backreaction in the form found in Shibata et al (1992). These imply solving two additional Poisson equations, which is computationally easy due to the use of FFTs.

The properties of neutron star matter are described by the equation of state of Lattimer & Swesty (1991), which we handle in tabular form. In addition to the fundamental hydrodynamic quantities, density, momentum, and energy, we follow the time evolution of the electron density in the stellar gas. Energy loss and changes of the electron abundance due to the emission of neutrinos are taken into account by an elaborate "neutrino leakage scheme", which employs a careful calculation of the lepton number and energy source terms of all neutrino types (Janka, 1991; Ruffert, Janka, and Schäfer, in preparation). Neutrinos are produced via thermal processes and lepton captures onto baryons. Matter is rendered optically thick to neutrinos due to neutrino-nucleon scattering and absorption of neutrinos onto baryons.

We simulate the coalescence of two identical, cold neutron stars with a baryonic mass of $\approx 1.6\ M_\odot$ and a radius of $\approx 15$ km. Initially the stars are spherical and orbit around the common center of mass with a center-to-center distance of 42 km and an orbital period of about 2 ms. The initial distributions of density and electron concentration are given from a model of a cold neutron star in hydrostatic equilibrium. The temperature is then chosen in such a way that the thermal energy is set to roughly 3% of the degeneracy level. The initial velocity distribution is varied between the three models which we simulate. The orbital velocities of the coalescing neutron stars are prescribed according to the motions of point masses as computed from the quadrupole formula. Spins are added to the neutron stars to take into account rotations around their axes vertical to the orbital plane. These spins are different from model to model to describe the cases with spin vectors parallel and antiparallel to the vector of the orbital angular momentum, and the case with rigid rotation, respectively.

The orbit decays due to gravitational wave emission, and after one revolution the stars are so close that dynamical instability sets in. Within 1 ms they merge into a rapidly spinning ($P \approx 1$ ms), high-density body ($\rho \approx 10^{14}$ g/cm$^3$) with a surrounding thick disk of material with densities $\rho \approx 10^{10} - 10^{12}$ g/cm$^3$ and rotational velocities of 0.3–0.5 c. Because of its high mass of $\approx 3\ M_\odot$ we expect the central object to collapse into a black hole essentially instantaneously.

The use of a "realistic" EOS in our simulations allows us to determine the temperature distribution in the merging stars and means that bulk viscosity effects of the matter are taken into account. Initially the neutron stars are cool and at a separation of several neutron star radii tidal deformations are insignificant. The neutron stars do not heat up noticeably even during the final stages of inspiral, since shear viscosity is small and the tidal deformations tend to increase the volume. The parts of the neutron stars closest to each other therefore experience tidal stretching and show cooling due to expansion rather than heating. Not earlier as when the stars start to merge, shearing effects of matter with opposite velocity components and local compression lead to a temperature increase. After merging we find peak temperatures of 30–40 MeV in two distinct "spots" inside the central, massive object, while the surrounding disk gas (which is only a few tenths of a solar mass) remains at temperatures of 4-8 MeV.

The peak emission of gravitational waves is short but powerful. A peak luminosity of a few times $10^{55}$ erg/s is reached for about 1 ms. The amplitudes of the gravitational waves reach $3 \cdot 10^{-23}$ for an assumed distance of 1 Gpc, and the typical frequency is near the dynamical value, 2 kHz.

Combining information about the fluxes of neutrinos and antineutrinos radiated into different directions, we are able to evaluate our models for the energy deposition by annihilation of $\nu\bar{\nu}$ into $e^+e^-$-pairs in the vicinity of the merger. This mechanism was suggested as a possible energy source for $\gamma$-ray bursts (see e.g., Mészáros & Rees, 1992; Mochkovitch et al, 1993). Although the final, dynamical phase of the coalescence happens within approximately 3 ms, the individual neutrino luminosities saturate at values of several $10^{52}$ erg/s only on the longer timescale of about 8 ms which the outer layers of the merged object need to reach their maximum temperatures (partly by shock heating). The integral rate of energy deposition by neutrino annihilation gets to a few $10^{50}$ erg/s during the phase of strongest neutrino emission. This rate has to be compared with the typical energy of $10^{50}$–$10^{51}$ erg observed in $\gamma$-ray wavelengths from bursts at cosmological distances. In order to yield this energy the merged object would have to radiate neutrinos with high luminosities for a period of 1–10 s. This is difficult to conceive in the considered scenario, since the merged object with a mass of around 3 $M_\odot$ will not escape the collapse to a black hole on a dynamical timescale of $\approx$ 1 ms. If one takes into account gravitational redshift effects and an efficiency of less than 100% for the conversion of $e^+e^-$-pair energy into $\gamma$-rays, our result is off from the desired value by more than a factor of 100. We therefore conclude that some other mechanism than $\nu\bar{\nu}$-annihilation has to be considered if $\gamma$-ray bursters are to be identified with coalescing neutron stars.

**References**

Blanchet, L., Damour, T., Schäfer, G., 1990, MNRAS, 242, 289.
Colella, P., Woodward, P.R., 1984, JCP, 54, 174.
Janka, H.-T., 1991, Dissertation, MPA report 587.
Lattimer, J.M., Swesty, F.D., 1991, Nucl. Phys., A535, 331.
Mészáros, P., Rees, M.J., 1992, ApJ, 397, 570.
Mochkovitch, R., Hernanz, M., Isern, J., Martin, X., 1993, Nature, 361, 236.
Shibata, M., Nakamura, T., Oohara, K., 1992, Prog. Theor. Phys., 88, 1097.